The simultaneous occurrence of superparamagnetism and superconductivity in nominal single crystalline $(Ca_{1-x}Pr_x)Fe_2As_2$ with $0 \leq x \leq 0.13$

B. Lv, F. Y. Wei, L. Z. Deng, Y. Y. Xue, and C. W. Chu*

Texas Center for Superconductivity and Department of Physics, University of Houston, Houston, Texas 77204-5002, USA

*Lawrence Berkeley National Laboratory, 1 Cyclotron Road, Berkeley, California 94720, USA

Abstract

The observation of non-bulk superconductivity with an unexpectedly high onset transition temperature $T_c$ up to ~ 49 K in non-superconducting single crystalline $CaFe_2As_2$ upon rare-earth doping has raised interesting questions concerning its origin. Several possibilities including interfacial mechanism have been proposed to account for the unusual observation. In an attempt to differentiate such propositions, we have carried out a systematic compositional and magnetic study on single crystals of nominal $(Ca_{1-x}Pr_x)Fe_2As_2$ with $0 \leq x \leq 0.13$ throughout the solubility range of Pr, which covers both the non-superconducting and superconducting regions. We found the unusual simultaneous occurrence of superparamagnetism and superconductivity with an x-independent $T_c$ and a close correlation of the superconducting volume fraction with the magnetic cluster density and As-defect density. The finding demonstrates a close relationship among superconductivity, superparamagnetism, and defects, consistent with the previously proposed interface-mechanism, and offers a possible future path to higher $T_c$.

The Fe-based layered pnictides and chalcogenides upon doping or under pressure constitute an interesting superconductor class with a transition temperature $T_c$ as high as 57 K [1], second only to that of ≤ 134 K at ambient [2] or 164 K under pressure [3] of the layered cuprate class. The Fe-based pnictide and chalcogenide superconductors comprise four families with their respective maximum $T_c$s at ambient as: 1111 (RFeAsO)—57 K [4]; 122 ($AEFe_2As_2$ or AE122)—38 K [5]; 111 (AFeAs)—18 K [6]; and 11 (FeSe)—10 K [7], where R = rare-earth, AE = alkaline earth, and A = alkaline, respectively. While doping by K or Na in the AE-sites generates bulk superconductivity with a maximum $T_c$ ~ 33–38 K as evidenced calorimetrically in all AE122 [8], pressure induces superconductivity with a similar maximum $T_c$ only in Ba122 and Sr122 [9]. Ca122 appears to behave differently from other 122 members. It becomes superconducting under quasi-hydrostatic pressure, but only below 12 K [10] and not under hydrostatic pressure [11]. It also exhibits a complex phase diagram with an unusual first-order tetragonal collapsed phase transition below ~ 160 K in the presence of pressure or on doping [12]. The most unusual of all observations is the non-bulk superconductivity with an onset $T_c$ up to 49 K in single crystalline Ca122 doped with rare-earth R = La, Ce, Pr, or Nd [13,14]. Such a $T_c$ is higher than that of any doped 122 or any known compounds within the R-AE-Fe-As chemical system, and than that predicted for a phonon-mediated superconductor [15]. The non-bulk nature of the superconductivity observed raises the question whether it is associated with the tetragonal collapsed phase, or a minute impurity phase of oxygen-deficient 1111 or of Ca122 at a yet-to-be-determined doping level. Subsequent studies show that the 49 K superconductivity cannot be associated with the tetragonal collapsed phase, since it appears also in samples without the tetragonal collapsed phase transition [13,14], nor can it be caused by a minute inclusion of oxygen-deficient 1111, since no superconductivity has been detected in Ca122 doped with Sm or Gd prepared under the same synthesis conditions, nor by a minute inclusion of optimally doped Ca122, because $T_c$ has been found to be independent of the Pr-doping [16,17]. More recent magnetic studies demonstrate [17] that the R-doped Ca122 single crystals consist of domains of weakly coupled Josephson-junction arrays (JJAs) with magnetic anisotropy up to ~ 200. The non-bulk superconductivity with a $T_c$ ~ 49 K has therefore been attributed to a possible interfacial $T_c$-enhancement in thin platelets of R-doped Ca122. The existence of the proposed superconducting domains in an otherwise chemically homogeneous R-doped Ca122 single crystal is indeed unexpected. To determine the basis for such unusual superconducting domains, we have therefore carried out a systematic compositional and magnetic investigation in single crystals of nominal $(Ca_{1-x}Pr_x)Fe_2As_2$ with $0 \leq x \leq 0.13$ throughout the solubility range of Pr, which covers both the non-superconducting and superconducting regions. Here we report the observations of the unusual simultaneous occurrence of superparamagnetism with superconductivity, which displayed two x-independent superconducting transitions, $T_{c1}$~ 49 K and $T_{c2}$ ~ 24 K, and a close correlation of the superconducting volume with the magnetic cluster density and the As-defect density. The observations demonstrate a close relationship among superconductivity, superparamagnetism, and defects. They are consistent with the previously proposed interface-mechanism for the enhanced-$T_c$ detected and offer a possible path to higher $T_c$s in the future.

The single crystals of $(Ca_{1-x}Pr_x)Fe_{2+y}As_{2-z}$ with $0 \leq x \leq 0.13$ investigated were grown in a reduced Ar-atmosphere sealed inside a silica tube by the self-flux technique as previously described [13]. Both the single crystal and powder X-ray diffraction patterns of the samples were obtained using a Rigaku DMAX III-B diffractometer and a Siemens SMART APEX diffractometer. The chemical analyses were performed using WDS on a JEOL JXA-8600 electron microprobe analyzer with 1 μm spot size giving an estimated systematic deviation below 0.5%. The magnetic measurements were carried out employing the 5 T Quantum Design Magnetic Property Measurement System.

The X-ray data show that the crystals studied are of high crystal quality without impurity phase to better than the resolution of 5%. The patterns are indexed nicely by a tetragonal cell with lattice parameters of $a = b \approx 3.91$ Å and $c \approx 11.5–11.8$ Å for x = 0–0.13. The WDS results of the single crystals throughout the Pr-solubility range of $0 \leq x \leq 0.13$ are shown in the ternary phase diagram in Fig. 1a. At least 5 well-separated spots are measured for each of the crystals and the statistical spread is below 0.5% represented by the symbols. The chemical homogeneity of the crystals is evident, consistent with the recent scanning tunneling microscopic study on the nominal $(Ca_{0.89}Pr_{0.11})Fe_2As_2$ [18]. Analyses of the WDS results, however, reveal a noticeable deviation from the 122-stoichiometric $(Ca_{1-x}Pr_x)Fe_2As_2$ (the blue star in Fig. 1a), represented by $(Ca_{1-x}Pr_x)_{1-y+z}Fe_{2+y}As_{2-z}$. By imposing the 5-atom-per-cell constraint, we find that all data points cluster around the line of $z \sim y > 0$. The compositions of the samples can therefore be rewritten as $(Ca_{1-x}Pr_x)Fe_{2+y}As_{2-y}$. The exact nature of such defects, e.g. antisite defects or Frenkel pairs, is yet to be determined by future study. However, for the convenience of later discussions, the defect density can be represented as y = 2(Fe-As)/(Fe+As). y is clearly influenced by the Pr-doping x. The x-dependence of y/2 is displayed in Fig. 1b for reasons that will be evident later. As x increases, y/2 first decreases from ~ 0.085 in our undoped Ca122 single crystal at x = 0 to ~ 0.005 at x ~ 0.07 and then increases linearly to ~ 0.035 at x = 0.13, the Pr-solubility limit. The non-monotonic variation of y with respect to x suggests that there may exist two different defect-structure types induced by Pr-doping below and above x ~ 0.06–0.07, respectively. As it will later be clear, x ~ 0.06–0.07 also defines the non-superconducting/superconducting boundary.

It has been shown that magnetic ordering is avoided in the iron-based superconductors through a delicate balance among the various Fe 3$d$-sub-bands. Net moments, therefore, are expected at or near the lattice defects that interrupt the Fe-As bonds and thus disturb the balance. Magnetic clusters around lattice vacancies were theoretically proposed [19]. To explore the effect of defects, the isothermal M-H loops were measured for all our as-synthesized crystals. The representative raw data of a crystal for Pr doping level x = 0.13 at 5, 15, 40, and 80 K are shown in Fig. 2a. A large anisotropy (M//$c$)/(M//$ab$) of ~ 7.5, e.g. at 5 K and 5 T, was detected, ruling out the possible origin of a randomly oriented magnetic inclusion for M. The raw data at both H-increase and H-decrease are included with no noticeable hysteresis above 0.1 T, implying a blocking temperature far lower than 5 K. Although at lower fields < 0.1 T and temperatures << $T_c$ a small remnant moment, e.g. ~ 50 emu/mol for this crystal, due to superconductivity, is observed, it is less than a few percent of the saturation moment and thus should not affect later analyses. Negligible relaxation is noticeable down to 2 K over our sequential measurements. As a result, the moments determined can be regarded as the equilibrium property of the system. This therefore allows us to carry out meaningful analyses of the data.

The magnetic data were analyzed in terms of the modified Langevin function M = $n\mu\mu_B$[1/tanh($z$)-1/$z$]+H·$\chi_{Pr}$ with $z = \mu\mu_B H/k_B(T+T_0)$, where $\mu_B$ and $k_B$ are Bohr magneton and Boltzmann constant, respectively, together with the fitting parameters $n$, $\mu$, and $T_0$ being the magnetic cluster concentration, the cluster magnetic moment, and an effective Curie-Weiss tempearture for the magnetic system, respectively [20]. The H·$\chi_{Pr}$ term is the contribution from the $Pr^{3+}$-ions, typically less than one third of the M. To avoid the fitting instability, an approximation of H·$\chi_{Pr}$ = x·H $\frac{N(3.5\mu_B)^2}{3k_B(T+T_{CW})}$ is used, where N is the Avogadro's number and $T_{CW}$ the Curie-Weiss temperature of the Pr-spins. Using the commonly accepted $Pr^{3+}$-moment of 3.5 $\mu_B$, the modified Longevin function fits well all M-H data for different x s at different temperatures with an x-independent $\mu$ = 8±2 $\mu_B$ and an x-dependent n and $T_o$, as exemplified in Fig. 2a. $T_o$ is found to be ~ + 80 (70) K at x = 0.04 , decreases to 0(1) K at x = 0.06, and becomes constant at ~ - 2(1) K for x > 0.06, implying the evolution of intercluster interaction from antiferromagnetic to weak ferromagnetic. To verify the evolution, the initial susceptibility $\chi_0 = \left.\frac{dM}{dH}\right|_{H=0}$ is extracted from the raw M(H,T) and displayed in Fig. 2b as a function of 1/T for different x s. From the temperature dependence of $\chi_0$, the intercluster interaction varies from antiferromagnetic ordering at ~ 12 K for x = 0.04 to non-interacting for x = 0.06 to slightly ferromagnetic for x = 0.13, in general agreement with the extracted $T_o$-x relation mentioned above.

It was previously reported that the Pr-doped Ca122 at x = 0.13 exhibits two distinct superconducting transitions with different field effects, one with an onset $T_{c1}$ at 49 K and the other, $T_{c2}$, at 24 K [13]. Later doping study showed that superconductivity appears at x ~ 0.06, and with a superconducting volume fraction $f \equiv 4\pi\chi_{dc}^{ab}$ measured at 5 K that grows continuously with x to x = 0.13 and with two clear transitions [16]. The low-temperature transition with an onset $T_{c2}$ ~ 24–29 K emerges abruptly at x ~ 0.06 and throughout the doping range 0.06 ≤ x ≤ 0.13, while the high-temperature transition with an onset $T_{c1}$ of 46–49 K appears suddenly at x ~ 0.08 and throughout the doping range of 0.08 ≤ x ≤ 0.13 (gray bands in Fig. 1b). The x-independence of the two $T_c$s suggests that the superconductivity observed cannot be caused by the variation of charge carriers through Pr-doping in the conventional sense, in agreement with the previous report [17]. Figure 3a shows the surprising correlation of *f* with *n*. Both *f* and *n* are zero for the nonsuperconducting samples with x < 0.06 and then scale with each other when superconductivity appears for samples with 0.06 < x ≤ 0.13, exposing a close relationship between superparamagnetism and superconductivity. To determine if such a close *f-n* relationship is caused directly by charge carrier concentration change via doping, as may be the case since *n* and x are related as shown earlier, and to explore the nature of the defects that induce the magnetic clusters, a systematic study on samples with different x s for different annealing times up to 200 hr at temperatures from 200 to 920 °C was made. Annealing at high temperature above 500 °C apparently suppresses both *f* and *n* rapidly with annealing time while keeping x unchanged for all samples tested, as shown in Fig. 3b. The observation demonstrates that the superconductivity in Pr-doped Ca122 is not a direct result of Pr-doping, consistent with the x-independent $T_c$ observed previously. However, annealing at 200 °C up to 100 hr shows no noticeable change in either *f* or *n*. The high annealing temperature and the long annealing time required to suppress *f* and *n* suggest the existence of rather large potential barriers for the defects. It is revealing to note that, in the superconducting region, *n* ~ y/2. It implies that the magnetic clusters form from special defect clusters, in agreement with a large potential barrier of the defects and the conjecture of theories for the occurrence of magnetism.

In conclusion, we have carried out systematic chemical and magnetic investigations on single crystals of $(Ca_{1-x}Pr_x)Fe_{2+y}As_{2-y}$ for 0 ≤ x ≤ 0.13. We observed the unusual simultaneous appearance of superparamagnetism and superconductivity with high $T_c$ in these samples. The magnetic cluster density scales with the superconducting signal size throughout the superconducting region of 0.06 < x ≤ 0.13 but is 0 for x ≤ 0.06, while the $T_c$ stays almost unchanged. At the same time, the magnetic cluster density is roughly equal to one-half the defect density, implying the importance of broken Fe-As bonds for magnetism. Prolonged annealing only at relatively high temperatures results in the suppression of the superconducting signal and the defect density without varying x, suggesting that the occurrence of superconductivity is related to defects with a relatively large potential barrier and not the direct result of carrier concentration change via Pr-doping. The above observations are consistent with the interface mechanism for enhanced-$T_c$ proposed. In view of its potential implication in further $T_c$-enhancement, microscopic investigations to unravel the working of the interfaces are warranted.

Acknowledgments

The work in Houston is supported in part by U.S. Air Force Office of Scientific Research contract FA9550-09-1-0656, the T. L. L. Temple Foundation, the John J. and Rebecca Moores Endowment, and the State of Texas through the Texas Center for Superconductivity at the University of Houston.

References

[1] Y. Kamihara, T. Watanabe, M. Hirano, and H. Hosono, J. Am. Chem. Soc. **130**, 3296 (2008).
[2] A. Schilling, M. Cantoni, J. D. Guo, and H. R. Ott, Nature **363**, 56 (1993).
[3] L. Gao, Y. Y. Xue, F. Chen, Q. Xiong, R. L. Meng, D. Ramirez, C. W. Chu, J. H. Eggert, and H. K. Mao, Phys. Rev. B **50**, 4260(R) (1994).
[4] Z. A. Ren et al., Chin. Phys. Lett. **25**, 2215 (2008); Z. Wei, H. Li, W. L. Hong, Z. Lv, H. Wu, X. Guo, and K. Ruan, J. Supercond. Nov. Magn. **21**, 213 (2008).
[5] M. Rotter, M. Tegel, and D. Johrendt, Phys. Rev. Lett. **101**, 107006 (2008); K. Sasmal, B. Lv, B. Lorenz, A. M. Guloy, F. Chen, Y. Y. Xue, and C. W. Chu, Phys. Rev. Lett. **101**, 107007 (2008).
[6] J. H. Tapp, Z. Tang, B. Lv, K. Sasmal, B. Lorenz, C. W. Chu, and A. M. Guloy, Phys. Rev. B **78**, 060505(R) (2008); X. C. Wang, Q. Q. Liu, Y. X. Lv, W. B. Gao, L. X. Yang, R. C. Yu, F. Y. Li, and C. Q. Jin, Solid State Commun., 148, 538 (2008).
[7] F. C. Hsu et al., Proc. Natl. Acad. Sci. USA **105**, 14262 (2008).
[8] K. Zhao, Q. Q. Liu, X. C. Wang, Z. Deng, Y. X. Lv, J. L. Zhu, F. Y. Li, and C. Q. Jin, J. Phys.: Condens. Matter **22**, 222203 (2010); F. Y. Wei, B. Lv, Y. Y. Xue, and C. W. Chu, Phys. Rev. B **84**, 064508 (2011).
[9] P. L. Alireza, Y. T. C. Ko, J. Gillett, C. M. Petrone, J. M. Cole, G. G. Lonzarich, and S. E. Sebastian, J. Phys.: Condens. Matter **21**, 012208 (2009).
[10] M. S. Torikachvili, S. L. Bud'ko, N. Ni, and P. C. Canfield, Phys. Rev. Lett. **101**, 057006 (2008); Y. Zheng, Y. Wang, B. Lv, C. W. Chu, and R. Lortz, New J. Phys. **14**, 053034 (2012).
[11] W. Yu, A. A. Aczel, T. J. Williams, S. L. Bud'ko, N. Ni, P. C. Canfield, and G. M. Luke, Phys. Rev. B **79**, 020511(R) (2009).
[12] A. Kreyssig et al., Phys. Rev. B **78**, 184517 (2008).
[13] B. Lv, L. Z. Deng, M. Gooch, F. Y. Wei, Y. Y. Sun, J. K. Meen, Y. Y. Xue, B. Lorenz, and C. W. Chu, Proc. Nat. Acad. Sci. USA **108**, 15705 (2011).
[14] S. R. Saha, N. P. Butch, T. Drye, J. Magill, S. Ziemak, K. Kirshenbaum, P. Y. Zavalij, J. W. Lynn, and J. Paglione, Phys. Rev. B **85**, 024525 (2012).
[15] W. L. McMillan, Phys. Rev. **167**, 331 (1968).
[16] C. W. Chu et al., J. Phys.: Conf. Ser. **449**, 012014 (2013).
[17] F. Y. Wei, B. Lv, L. Z. Deng, J. Meen, Y. Y. Xue, J. E. Hoffman, and C. W. Chu, unpublished.
[18] I. Zeljkovic, D. Huang, C. L. Song, B. Lv, C. W. Chu, and J. E. Hoffman, Phys. Rev. B **87**, 201108(R) (2013).
[19] K. W. Lee, V. Pardo, and W. E. Pickett, Phys. Rev. B **78**, 174502 (2008).
[20] For example, B. P. Krustalev, A. D. Balaev, and V. M. Sosnin, Solid State Commun. **95**, 271 (1995).

Figure Captions

FIG. 1. (a) The ternary phase diagram of the $(Ca_{1-x}Pr_x)$-Fe-As system based on ~ 40 raw WDS data points, where x varies between 0 and 0.13. The blue star represents the stoichiometric $CaFe_2As_2$. (b) The evolution of the superparamagntic clusters density $n$ (red inverted triangles) and the As-deficiency y (black triangles) with the Pr doping level x. The error bars represent the data spreads over 5 well separated measurement points. The dashed line represents the superconducting and non-superconducting (NSC) boundary. The gray bands represent the onset of two superconducting transitions $T_{c1}$ and $T_{c2}$.

FIG. 2. (a) The isothermal M-H loops for the x = 0.13 crystal and the Langevin fit (solid lines) at different temperatures. (b) The initial *dc* susceptibility $dM/dH|_{H=0}$ normalized at 2 K and up to 50 K for different crystals with: x ≈ 0.13 and n ≈ 0.03 (red triangles); x ≈ 0.06 and n ≈ 0.006 (green circles); and x ≈ 0.04 and n ≈ 0.004 (blue squares). The straight black dashed line represents the $c \propto 1/T$ Curie-Weiss fit.

FIG. 3. (a) The correlation of superconducting volume fraction $f$ with superparamagntic clusters density $n$ for preannealed crystals with different doping levels. (b) Annealing effects on the normalized superconducting volume fraction $f$ for different crystals.

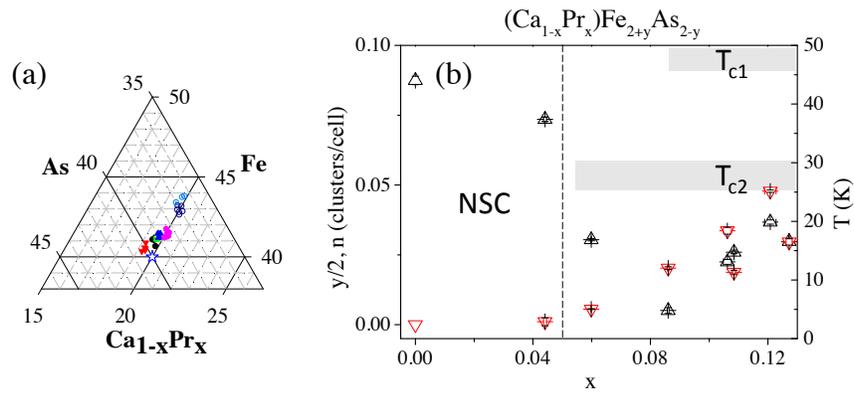

Fig. 2

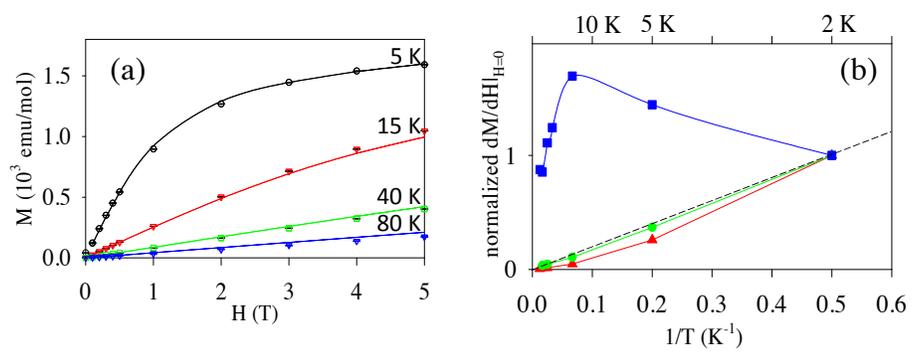

Fig. 3

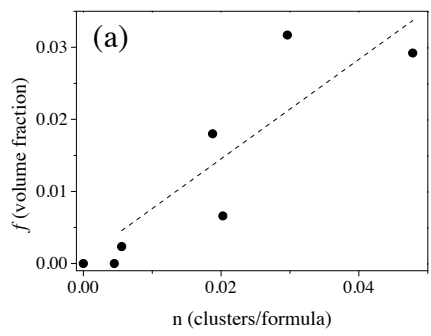 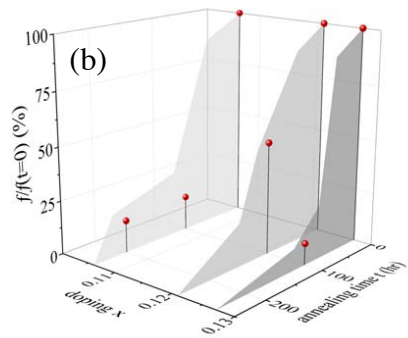